\documentstyle [twocolumn,prb,aps] {revtex}
\begin{document}
\draft
\begin{title}
{Comment on ``Density functional theory study of some structural \\
and energetic properties of small 
lithium clusters''  [J. Chem. Phys. 105, 9933 (1996)]}
\end{title} 
\author{Constantine Yannouleas and Uzi Landman }
\address{
School of Physics, Georgia Institute of Technology,
Atlanta, Georgia 30332-0430 }
\date{January 1997}
\maketitle
\narrowtext
The experimentally determined size-evolutionary patterns (SEPs) of simple 
metal clusters pertaining to ionization potentials (IPs), 
electron affinities, and
monomer separation energies have attracted much attention in the past ten
years \cite{dehe,brec}, since they may reflect the electronic
shell structure of clusters (the major features of SEPs are associated 
with steps at magic numbers, but often there is in addition substantial fine
structure, such as odd-even alternations). 

Due to the required computational effort, first-principles (FP) theoretical 
studies (which incorporate the ionic geometry) of such SEPs are usually 
limited to cluster sizes of the order of ten atoms. \cite{bona,barn} 
Recently, however, systematic theoretical investigations of such SEPs have
been performed for a broad range of cluster sizes using the jellium-related 
Shell-Correction-Method (SCM) approach. \cite{yann1,yann2,yann3,yann4} 

In a recent article, Gardet {\it et al.\/} \cite{gard} have studied the IPs 
of small lithium clusters Li$_N$ (with $N \leq 12$) using a FP -
density functional theory (DFT) and found reasonable agreement between 
theory and experiment. Furthermore, through a comparison of their results
with those obtained from Kohn-Sham local-density-approximation (KS-LDA)
calculations on a {\it spherical\/} jellium background, \cite{ekar,pusk}
they concluded that the jellium model is inadequate for 
a proper description of the IPs of such systems.

The purpose of this comment is to clarify that, while modelling the ions
in a cluster via a uniform jellium background is certainly an approximation,
the above conclusion of Gardet {\it et al.\/} pertains to limitations 
introduced through neglect of deviations from spherical symmetry,
rather than to the jellium approximation itself.
Indeed, in a series of papers, \cite{yann2,yann3,yann4}
we have demonstrated that consideration of triaxial
shape deformations drastically improves the agreement between the 
jellium approximation and experiment for all instances of the aforementioned
SEPs and for sizes up to 100 atoms, as well as for a variety of
metal species (namely, alkali metals, such and Na and K, and noble metals,
such as Cu and Ag). 

To further elucidate the importance of shape deformations, we display in
Fig.\ 1 the IPs of small Li$_N$ clusters (in the same size-range as with
Ref.\ \onlinecite{gard}) calculated \cite{note} with our SCM for
three different families of shapes of the jellium background 
[namely, spherical, spheroidal (axially symmetric), and ellipsoidal 
(triaxial)], and compare them to the experimental 
measurements. \cite{broy} 

Fig.\ 1 reveals that spherical shapes (top panel) exhibit a characteristic
sawtoothed profile, well known from previous spherical KS-LDA
studies \cite{ekar} and similar to the
curve \cite{vezi} labeled jellium-LDA in Fig.\ 12 of Ref.\ \onlinecite{gard}.
Apart from major-shell closures, this sawtoothed profile describes the data 
rather poorly (notice in particular the absence of fine structure between 
major shell closures at $N=2$, 8 and 20).

The spheroidal model (middle panel) exhibits substantial improvement
in describing the experimental trend. Furthermore, the ellipsoidal case
(bottom panel) improves the agreement between the SCM results and experiment
even further, in particular in the size range $11 \leq N \leq 14$. 
The essential improvement introduced by the deformations over the spherical 
case concerns the very good description of the subshell closure at $N=14$ and 
of odd-even alternations between major shell closures.

In summary, we have illustrated once again that, 
within the jellium approximation,
deformed cluster shapes provide an adequate description of the observed
systematic size dependence of the properties of simple metal clusters
and should necessarily be employed in comparisons with other theoretical 
approaches.

This research is supported by the US Department of Energy (Grant No.
FG05-86ER-45234). Studies were performed at the 
Georgia Institute of Technology Center for Computational Materials Science.

\begin{figure}
\caption{IPs for Li$_N$ clusters in the range $2 \leq N \leq 16$.
Solid dots: Theoretical results derived from the SCM (see Refs.\ 
\protect\onlinecite{yann2,yann3}) using three different shape models for
the jellium background. Top panel: The spherical model 
compared to experimental data (open squares). 
Middle panel: The spheroidal model compared to experimental data. 
Lower panel: The ellipsoidal model compared to 
experimental data. The experimental measurements (open squares) were taken 
from Ref.\ \protect\onlinecite{broy}.}
\end{figure}

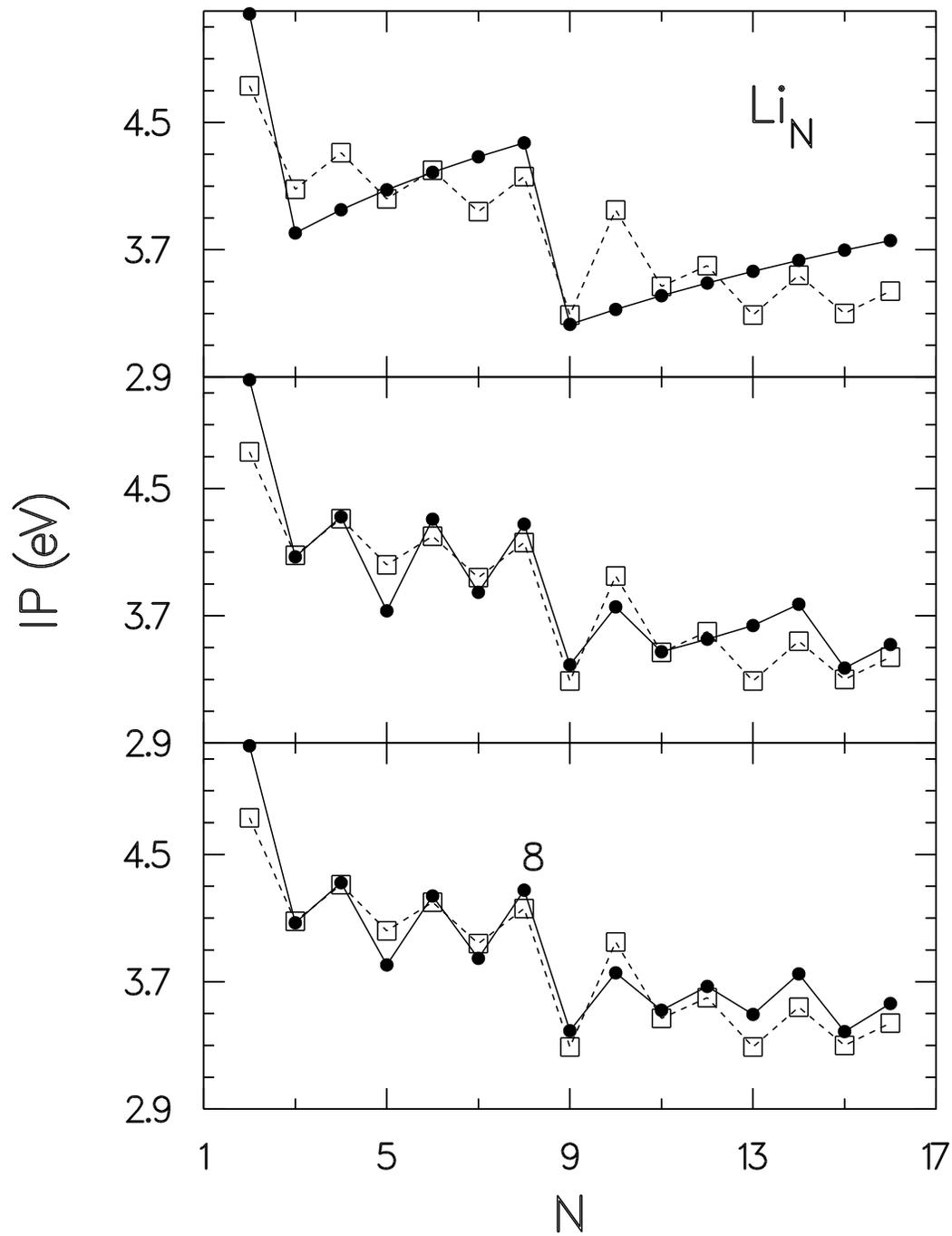
\end{document}